\documentclass[]{aa}
\usepackage{graphicx}
\usepackage{url}
\usepackage{txfonts}
\usepackage[]{hyperref}
\usepackage[]{natbib}
\usepackage[]{twoopt}
\bibpunct{(}{)}{;}{a}{}{}

\makeatletter
\newcommandtwoopt{\citeads}[3][][]{\href{http://adsabs.harvard.edu/abs/#3}%
{\def\hyper@linkstart##1##2{}%
\let\hyper@linkend\@empty\citealp[#1][#2]{#3}}}
\newcommandtwoopt{\citepads}[3][][]{\href{http://adsabs.harvard.edu/abs/#3}%
{\def\hyper@linkstart##1##2{}%
\let\hyper@linkend\@empty\citep[#1][#2]{#3}}}
\newcommandtwoopt{\citetads}[3][][]{\href{http://adsabs.harvard.edu/abs/#3}%
{\def\hyper@linkstart##1##2{}%
\let\hyper@linkend\@empty\citet[#1][#2]{#3}}}
\newcommandtwoopt{\citeyearads}[3][][]%
{\href{http://adsabs.harvard.edu/abs/#3}
{\def\hyper@linkstart##1##2{}%
\let\hyper@linkend\@empty\citeyear[#1][#2]{#3}}}
\makeatother

\begin{document}

\defcitealias{1999SoPh..184..421N}{N99}
\defcitealias{1973apds.book.....D}{DNR73}
\defcitealias{Pierce+Breckinridge.1973}{PB73}
\defcitealias{1984sfat.book.....K}{KFBT84}
\defcitealias{1984SoPh...90..205N}{NL84}

\title{How different are the Li\`ege and Hamburg\\ atlases
  of the solar spectrum?}

\author{H.-P. Doerr\inst{1}
  \and N. Vitas\inst{2,3}
  \and D. Fabbian\inst{1,2,3}
}
%\offprints{}

\titlerunning{How different are atlases of the solar spectrum?}

\institute{%
  Max-Planck-Institut f\"ur Sonnensytemforschung (MPS),
  %Justus-von-Liebig-Weg 3, 
  37077 G\"ottingen, Germany
  \and
  Instituto de Astrof\'isica de Canarias (IAC),
  %Calle V\'ia L\'actea s/n, 
  38205 La Laguna (Tenerife), Spain
  \and
  Departamento de Astrof\'isica, Universidad de La Laguna (ULL),
  38206 La Laguna (Tenerife), Spain
}

\date{\today}

\abstract
% context heading (optional)
{%
  The high-fidelity solar spectral atlas prepared by Delbouille and
  co-workers (Li\`ege atlas), and the one by Neckel and co-workers (Hamburg atlas),
  are widely recognised as the most important reference spectra of the Sun
  at disc-centre in the visible wavelength range. Both datasets serve as
  fundamental resources for many researchers, in particular for chemical
  abundance analysis. But despite their similar published specifications (spectral
  resolution, noise level), the shapes of spectral lines in the two atlases
  differ significantly and systematically.
}
% aims
{%
  Knowledge of any instrumental degradations is imperative to fully exploit
  the information content of spectroscopic data. We seek to investigate the
  magnitude---and to explain the possible sources---of these differences. We
  provide the wavelength-dependent correction parameters that need to be
  taken into account when the spectra are to be compared with
  e.~g. synthetic data.
}
% methods
{%
  A parametrically degraded version of the Hamburg spectrum is fitted to the
  Li\`ege spectrum. The parameters of the model (wavelength shift,
  broadening, intensity scaling, intensity offset) represent the different
  characteristics of the respective instruments, observational strategies
  and data processing.
}
% results heading (mandatory)-
{%
  The wavelength scales of the Li\`ege and Hamburg atlases differ on average
  by 0.5\,m\AA{} with a standard deviation of $\pm 2$\,m\AA{}, except for a
  peculiar region around 5500\,\AA. The continuum levels are offset by up to
  18\% below 5000\,\AA{} but stably at a~0.8\% difference towards the red.
  We find no evidence for spectral straylight in the Li\`ege spectrum. Its
  resolving power is almost independent of wavelength but limited to about
  216\,000, between two to six times less than specified. When accounting
  for the degradations determined in this work, the spectra of the two
  atlases agree within a few parts in $10^{3}$.  }
{}
{}


\maketitle

\section{Introduction}

The solar spectrum is characterised by the presence of millions of spectral
features which are caused by atomic and molecular energy transitions. The
intensities and shapes of these spectral lines show a continuous variation
on small spatial scales and at short temporal scales. However, in the time-
and space-averaged visible and near-infrared spectrum of the Sun, only the
lines formed in the chromosphere show some variation with the solar cycle
while photospheric lines remain unchanged---with the exception of the
\ion{Mn}I\,5394.7\,\AA{} line
\citepads{1987ApJ...314..808L,1988Sci...240.1765L,2007ApJ...657.1137L}.

Because of this lack of variability, it makes sense to record the solar
spectrum with high fidelity (spectral resolving power
$\lambda/\delta\lambda\sim$\,500\,000, signal-to-noise (S/N) ratio
$\gtrsim 1000$) for a wide range of wavelengths, tabulating it in the form
of an atlas so that it can be used as a standard for various research
fields. Atlases of the spectrum of a spatially averaged region at the centre
of the solar disc (intensity or photospheric spectrum) are used as a
benchmark for simulations
\citepads[e.g.][]{2013A&A...554A.118P,2015ApJ...802...96F}, for the
derivation of solar chemical abundances
\citepads[e.g][]{2008A&A...488.1031C,2009ARA&A..47..481A}, and as a
reference for spectroscopic data calibration
\citepads[e.g.][]{2004A&A...423.1109A,2011A&A...535A.129B}. Likewise,
atlases of the disc-averaged sunlight (flux spectrum) play a key role as a
solar standard in stellar astrophysics, for example in the derivation of
relative chemical abundances.

The first high-quality atlas of the solar disc-centre spectrum was prepared
by \citetads{1940pass.book.....M} in Utrecht (``\textit{Photometric atlas of
  the solar spectrum from $\lambda$\,3612 to $\lambda$\,8771}'', commonly
referred to as the Utrecht atlas) using observations carried out at the
Mt. Wilson observatory. Since then, a variety of further high-fidelity
spectral atlases of the Sun have been prepared based on steadily improving
instrumentation, mainly at the Jungfraujoch observatory in the Swiss Alps
\citepads[see the review by][]{1995ASPC...81...32D}, and in the United
States at the Sacramento Peak observatory \citepads{1976hrsa.book.....B},
and at the Kitt Peak National Solar Observatory (NSO) \citepads[see,
e.\,g. the review by][]{1995ASPC...81...66H}. All these developments
culminated with the Fourier-Transform Spectrometer (FTS) of the Kitt Peak
McMath-Pierce solar facility in the early 1980s.

Today's solar observations focus on imaging spectro-polarimetry with high
spatial resolution, often at the expense of spectral coverage and spectral
purity.  Such observations of narrow spectral ranges can be used to target
specific spectral lines, e.\,g. for chemical abundance analysis
\citepads[e.g.][]{1994A&AS..104...23K,2009A&A...507..417P,2015A&A...579A..88C},
but many of the applications mentioned above rely instead on a broad-band
high-quality reference spectrum.

Three different high-fidelity atlases of the intensity spectrum of the Sun
in the visible light are commonly used today: 1) The ``\textit{Photometric
  Atlas of The Solar Spectrum from $\lambda$\,3000 to $\lambda$\,10000}''
which was prepared with data from the Jungfraujoch Observatory, and which
was first published in a printed edition by the University of Li\`ege,
Belgium \citepads[][hereafter
\citetalias{1973apds.book.....D}]{1973apds.book.....D}. We refer to this
atlas as the Li\`ege visible light disc-centre atlas, or just Li\`ege atlas
for brevity; 2) the ``\textit{Spectral Atlas of Solar Absolute Disk-Center
  Intensity from 3290 to 12510\,\AA}'', which was prepared using data from
the FTS of the McMath-Pierce telescope and which was published in digital
format by \citetads{1999SoPh..184..421N}, hereafter
\citetalias{1999SoPh..184..421N}, at the University of
Hamburg, Germany. We refer to this atlas as the Hamburg disc-centre atlas,
or just Hamburg atlas; 3) ``\textit{An atlas of the spectrum of the solar
  photosphere from 3570 to 7405\,\AA}'' \citepads{1998assp.book.....W} which
was also made with data from the McMath FTS. We focus on the Li\`ege and
Hamburg atlases as these are the most widely used ones. Also, the Wallace
atlas has a more limited wavelength coverage and an obviously inferior
quality when compared to the Hamburg atlas
\citepads[e.g.][]{2008ApJ...686..731A}.

The Li\`ege and Hamburg atlases were made using very different instruments,
at different observing sites, and during different epochs along the solar
cycle. This naturally rises the question about their respective quality and
about how well they compare.
In fact, various authors find inconsistencies in the parameters retrieved
from these atlases. Several studies report the need for a careful
consideration of the continuum level of the Li\`ege atlas
\citepads[e.g.][]{1975A&A....45...19A,1984A&AS...55..143R}. \citetads{1982A&AS...47..193G}
compared equivalent widths derived from their observational data with those
from the Li\`ege atlas and find the latter ones to be larger by 5\,\% on
average.  \citetads{2000A&A...359..729A} argue that the Hamburg atlas is
superior over the Li\'ege one in terms of wavelength calibration and
continuum placement, based on the comparative study by
\citetads{1998A&AS..129...41A}. \citetads{2008A&A...488.1031C,2009A&A...498..877C,2011SoPh..268..255C}
find significantly different line shapes in the Hamburg and Li\`ege atlases
when comparing specific spectral lines. Those authors speculate it may be
caused by different levels of telluric line contamination and invoke the
necessity for a new high-quality solar atlas. \citetads{2015A&A...573A..26S}
report an excellent agreement of equivalent widths derived from these
atlases and their synthetic spectra.

To the best of our knowledge, so far no systematic study is available on the
compatibility of these two important spectral atlases, except for the
investigation of the wavelength calibration carried out by
\citetads{1998A&AS..129...41A}. We intend to provide a general, quantitative
measure of the differences between the Li\`ege and Hamburg atlases. A direct
comparison of basic line parameters such as their wavelength positions and
equivalent widths is not adequate for this purpose, as they are mostly
independent of the spectral resolution. On the other hand, absolute
parameters such as the line width and line depth are very sensitive to the
spectral resolution. Our approach does not depend on detailed knowledge of
the instrument specifications because we model the instrumental degradations
to find those that result in the best match between the Hamburg and Li\`ege
spectra.

\section{The spectral atlases}\label{sec:data}

We restrict the analysis to the aforementioned Li\`ege atlas and the Hamburg
atlases. A variety of further nicknames can be found in the literature for
these two atlases, e.g. Jungfraujoch atlas or Delbouille atlas for the
former and Neckel atlas, Kitt-Peak atlas, FTS atlas, and variations thereof
for the latter. In \citetalias{1999SoPh..184..421N}, it is stated that the
Hamburg atlas was prepared by ``Brault \& Neckel (1987)'', a reference that
was adopted by some researchers and which adds to the confusion as, to our
best knowledge, there is no corresponding publication. The naming is even
more unclear as a variety of further solar spectral atlases with similar
nicknames (e.g. FTS atlas) made with McMath FTS data were published by other
authors. In fact, we sometimes found references in the literature to one of
these atlases while obviously a different one was actually used.
In an attempt to lift some of that confusion, we briefly introduce the
various visible-light, disc-centre and disc-averaged McMath FTS atlases
before we proceed with a more detailed description of the Li\`ege and
Hamburg atlases.

The ``\textit{Solar flux atlas from 296 to 1300\,nm}'' is probably the most
widely used reference spectrum of the disc-averaged Sun in visible light. It
was originally published in a printed edition \citepads[][hereafter
\citetalias{1984sfat.book.....K}]{1984sfat.book.....K} but is also available
in digital format.%
\footnote{\url{ftp://vso.nso.edu/pub/atlas/fluxatl}; see
  also \url{http://kurucz.harvard.edu/sun/fluxatlas}}
This atlas is often referred to as the NSO atlas.

Kurucz published%
\footnote{\url{http://kurucz.harvard.edu/sun/fluxatlas2005/}}
a re-reduced version which he corrected for telluric lines
\citepads{2005MSAIS...8..189K}.  Another flux atlas---based on different
data---with a correction for telluric features was published%
\footnote{\url{ftp://vso.nso.edu/pub/Wallace_2011_solar_flux_atlas/}}
by \citetads{2011ApJS..195....6W}.

The first official disc-centre atlas made with McMath FTS data is
``\textit{An atlas of the spectrum of the solar photosphere from 13,500 to
  28,000\,cm$^{-1}$ (3570 to 7405 \AA{})}'' by
\citetads{1998assp.book.....W}%
\footnote{also availbe online: \url{ftp://vso.nso.edu/pub/atlas/visatl/}}
which was later extended to a wider wavelength range
\citepads{2007assp.book.....W}.

The aforementioned Hamburg disc-centre atlas was prepared from different
FTS observations than both of those prepared by Wallace et\,al., but there
is also a Hamburg flux atlas which is based on the same data as the
\citetalias{1984sfat.book.....K} flux atlas.

Recently, \citetads{2015A&A...573A..74S} published an atlas of the solar
spectrum at the disc-centre and at the limb. His disc-centre data is
identical with the Hamburg spectrum, except for a different continuum
normalisation \citep[][priv. comm.]{sten2015pc}.%
 \footnote{The continuum was re-normalised to the maximum intensity in each
   1000\,\AA{} wide interval.}
The limb spectrum stems from archival McMath FTS data recorded for a former
investigation \citepads{1983A&AS...54..505S}.

More atlases have been made with the McMath FTS, covering the near-infrared
and mid-infrared ranges, sunspot atmospheres etc. \citepads[see
e.g.][]{1995ASPC...81...66H}, but these shall not be discussed here.

\subsection{The Li\`ege disc-centre atlas}
\label{sec:liegeatlas}

The data for this atlas was recorded between 1973 and 1988 with a scanning
double-pass grating spectrometer at the Jungfraujoch scientific station in
the Swiss Alps. The observatory is located at an altitude of 3580\,m and
thus enables observations through an extremely dry atmosphere during a few
days in the year.

Details about the instrument, observational procedures and the data
processing can be found in the introduction to the atlas
\citepalias{1973apds.book.....D}, in the introduction to the older infrared
atlas made at Jungfraujoch \citepads{1963apss.book.....D}, and in the review
by \citetads{1995ASPC...81...32D}. A coelostat with mirrors of 30\,cm
diameter fed the fixed telescope which had a focal length of 1250\,cm. The
solar image of 11.5\,cm diameter was centred on the 2.5\,cm long
(400\arcsec{} projected on the Sun) entrance slit of a prism monochromator
that was used for order selection. The main spectrometer was based on a
Fastie-Ebert setup with 730\,cm focal length, modified for double-pass. Two
different gratings were used, one with~600 rules\,mm$^{-1}$ below
3586\,\AA{} and one with~300~rules\,mm$^{-1}$ above. The theoretical
spectral resolving power is specified with 1\,250\,000 and 500\,000 at the
blue and red ends of the atlas. The instrumental profile measured with a
laser is provided for four grating orders. The authors emphasise the
excellent straylight suppression and spectral purity obtained with the
double-pass configuration.

The atlas covers the range between 3000 and 10000\,\AA{} and was pieced
together from many individual segments. Each segment is only a few \AA{}
wide and was prepared from~50 repeated scans which were acquired within
approximately 45\,minutes. Detector dark currents were measured before and
after each segment was recorded and the mean was subtracted. The data of
each segment were processed with several analogue and digital data reduction
steps (binning, outlier-rejection, adaptive low-pass filtering) online
before the final spectrum was stored on magnetic tape.

The wavelength scale for each of the segments was derived from the line list
of \citet{Pierce+Breckinridge.1973}, hereafter
\citetalias{Pierce+Breckinridge.1973}.  The wavelengths are therefore given
in the solar frame of reference and are corrected for the radial velocities
of the observations but not for the gravitational redshift. The position of
the continuum was placed at 100\% of the intensity scale by interactively
straightening the individual segments to match with the slightly overlapping
adjacent segments. The noise level is not explicitly stated, but we measured
a very high S/N of~$\sim$\,6000 in the continuum at~6300\,\AA. The data for
the digital version of the atlas was resampled to constant wavelength
increments of~2\,m\AA.

The data was originally distributed on magnetic tapes and copies thereof
seem to be used at various research institutes. Unfortunately, the complete
atlas does not seem to be available for download. Individual spectral
regions can be selected interactively and downloaded in ASCII format from a
database provided by the Observatoire de
Paris.\footnote{\url{http://bass2000.obspm.fr}}

\subsection{The Hamburg disc-centre atlas}
\label{sec:hamburgatlas}

The data for this atlas was recorded around 1980%
\footnote{The exact
  observing dates are not published (not even the year). But the first data
  in the FTS archive reaches back to 1977 and a paper making use of the data
  \citepalias{1984SoPh...90..205N} was submitted in 1983.}
with the FTS at the McMath-Pierce solar telescope at the Kitt-Peak National
Observatory. The telescope is located at an altitude of 2096\,m.

\begin{table}
  \caption{Summary of the basic atlas properties.}             
  \label{table:1}   
  \centering                              
  \begin{tabular}{l c c}          
    \hline\hline
    Parameter & Li\`ege & Hamburg\\ 
    \hline
    Observatory & Jungfraujoch & Kitt Peak NSO \\
    (altitude)  & (3580\,m) & (2096\,m) \\
    Instrument & double-pass grating & FTS\\
    Year span & 1973--1988& $\ge$1977, $\le$1983\\
    Wavel. range & 3000--10000\,\AA& 3290--12510\,\AA\\
    ${\lambda}/{\delta\lambda}$ & $>$500\,000 ($\sim$\,216\,000*)& up to
                                                                   520\,000\\
    S/N at\,630\,nm & $\sim$\,6000* & $\sim$\,3000*\\
    avg. area & 400\arcsec{} slit & circular, 5\arcsec$^1$ \\
    comments & very dry site & abs. intensities \\
    \hline
  \end{tabular}
  \tablefoot{Values are as given by the authors of the
    atlases. Values marked with an
    asterisk were determined in this work.\\$^1$ Assumed in this work,
    actual value unknown but $\leq$10\arcsec.}
\end{table}

The disc-centre atlas and an associated flux atlas are a by-product of a
former investigation by \citetads{1984SoPh...90..205N}, hereafter
\citetalias{1984SoPh...90..205N}. The Hamburg flux atlas is based on the
very same FTS observations than the \citetalias{1984sfat.book.....K} atlas
for which details of the observational parameters and instrument settings
were published. Unfortunately, only few details about the disc-centre data
are published and so far we did not succeed in identifying the used FTS
scans in the data archive provided by NSO. The available information in
\citetalias{1984SoPh...90..205N} and \citetalias{1999SoPh..184..421N} however
indicates that instrumental settings and observational procedures were
similar for flux and disc-centre data.

Both atlases cover the wavelength range between~3290\,\AA{} and~12510\,\AA{}
and were pieced together from seven individually measured segments in
slightly overlapping wavelength intervals. The spectral windows were
selected with a low-dispersion monochromator and optical filters. Each
segment consists of several co-added FTS scans to increase the S/N
ratio. Each scan typically took about five minutes. The number of summed
scans for each segment of the disc-centre atlas is not published but for the
flux atlas it varies between 16 and 36, leading to total observing times of
around two hours \citepalias[Table~1 in][]{1984sfat.book.....K}. The exact
size of the mask used to define the averaging area on the solar disc is not
known. According to the meta-data provided by the FTS archive, the data for
Wallace' atlas was recorded with a circular mask of~5\arcsec{} diameter and
we may assume that similar settings apply for the observations used for the
Hamburg atlas. Other commonly used masks were circular with 10\arcsec{} or
rectangular with a similar effective area
\citepads[e.\,g.][]{1988A&AS...72..473B}.

According to \citetalias{1999SoPh..184..421N} and
\citetalias{1984SoPh...90..205N}, the wavenumber sampling $\delta\sigma$ was
set to values between 0.035\,cm$^{-1}$ at the blue end and 0.012\,cm$^{-1}$
at the red end. Assuming a strictly Nyquist-critical sampling,%
\footnote{This is not necessarily the case. Many FTS observations were
  acquired with at least three samples per resolution element
  \citepads[e.g.][]{1985hra..conf....3B}.}
that translates to a spectral resolving power $\sigma/2\delta\sigma$ of at
least 350\,000 for all scans in both atlases. Because of the equidistant
wavenumber sampling of a FTS, the resolving power changes with wavelength
and reaches values of up to 520\,000 at the blue end of some segments
(cf. Fig.~\ref{fig:jfjres}). The wavelength calibration for the disc-centre
atlas was derived from the wavelengths given in the line list of
\citetalias{Pierce+Breckinridge.1973}, just like in case of the Li\`ege
spectrum. \citetalias{1999SoPh..184..421N} states that the fractional
wavelength errors are smaller than~$1.5\times10^{-7}$.

The atlas provides the line spectrum and the position of the continuum, both
in intensity units (W\,cm$^{-2}$\,ster$^{-1}$\,\AA$^{-1}$) with an accurate
(<0.5\%) absolute calibration \citepalias{1984SoPh...90..205N} which is
based on the earlier measurements of \citetads{1967ZA.....65..133L}. In this
work, we exclusively use the Hamburg spectrum with the continuum normalised
to unity, that is by dividing the line spectrum by the continuum
intensities.%
\footnote{Interestingly, this results in some intensity values $>1.0$ with a
  maximum of $\sim$\,1.01. The median of the intensities $>1.0$ is
  $\sim$\,1.001, which is marginally compatible with the stated 0.5\%
  accuracy of the intensity values.}
The S/N ratio is not explicitly specified for the disc-centre atlas. We
measured a S/N of~$\sim$\,3000 in the continuum at ~6300\,\AA. The data has
a variable wavelength sampling, a consequence of the conversion from
wavenumber to wavelength units.

The Hamburg atlas is available in ASCII format from the ftp server of the
University of
Hamburg\footnote{\url{ftp://ftp.hs.uni-hamburg.de/pub/outgoing/FTS-Atlas/}}.

\subsection{Remarks}
%\vspace{1cm}

\begin{figure}[tbp]
  \resizebox{\hsize}{!}{\includegraphics{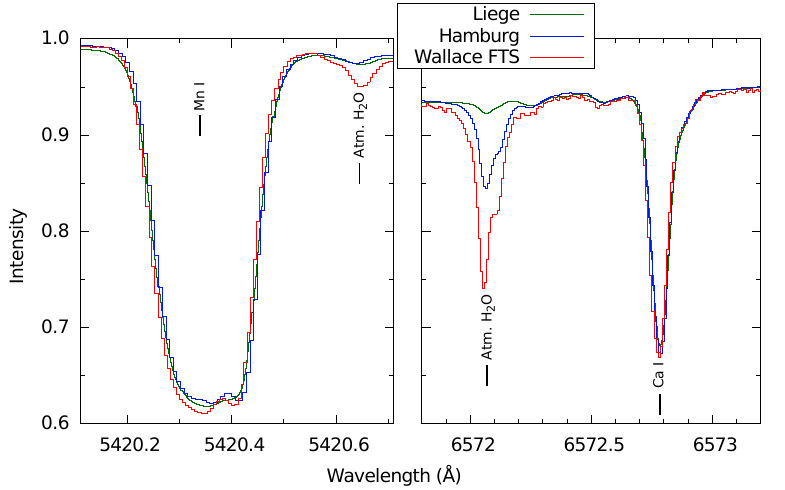}}
  \caption{Details in the Li\`ege, Hamburg and the Wallace disc-centre
    atlases in two spectral regions. Line identifications are from the list
    of \citetalias{Pierce+Breckinridge.1973}.}
  \label{fig:fidelity-compare}
\end{figure}

A summary of the basic specifications of both atlases is shown in
Table~\ref{table:1}. From these numbers the Li\`ege spectrum is expected to
be of higher quality than the Hamburg spectrum. A direct side-by-side plot
in Fig.~\ref{fig:fidelity-compare} (cf. also Fig.~\ref{fig:example.fit})
indicates a very similar
noise level, but it also reveals a loss of spectral detail in the Li\`ege
spectrum where the hyper-fine splitting of the manganese line
at~5420.4\,\AA{} can barely be seen while it is clearly resolved in the
Hamburg spectrum and also in Wallace' atlas. The right panel shows details
in the red wing of the H$\alpha$ line, a region that is heavily contaminated
with telluric H$_2$O lines. The H$_2$O features are much less pronounced in
the Li\`ege spectrum than in the other atlases.%
\footnote{Wallace' atlas is also provided with a correction for telluric
  lines. The correction sometimes fails in the core of very deep lines,
  though, therefore we used the non-modified spectrum here.}
The Wallace atlas shows a pronounced high-frequency fringe pattern with
varying amplitude. The behaviour mentioned above can be observed at all
wavelengths covered by the atlases and is not unique to the spectral windows
selected here. Despite these issues, we found that the Li\`ege and Hamburg
atlases provide a more consistent quality throughout the whole spectral
range. For this reasons, we discarded the Wallace atlas and restrict the
analysis to the Hamburg and Li\`ege spectra only \citepads[see
also][]{2008ApJ...686..731A}.

\section{Method}

We assume that the higher-resolved Hamburg spectrum ($\text{I}_\text{Ha}$)
can be transformed to the Li\`ege spectrum,
\begin{equation}
  \text{I}[\lambda_1,\lambda_2]_{\text{Li}}
  = S\left(\text{I}[\lambda_1+L,\lambda_2+L]_{\text{Ha}} * \Psi(B) +
    O\right),
  \label{eq:transformation}
\end{equation}
by modelling the degradations in the wavelength interval
$\lambda_1+\text{L}$ to $\lambda_2+\text{L}$ as an additional broadening
($B$) of the spectral lines due to a lower spectral resolution, a wavelength
shift ($L$) due to fluctuations in the wavelength calibrations, a scaling
factor ($S$) due to differently placed continuum levels, and an offset ($O$)
due to undispersed scattered light. The broadening is modelled as the
convolution with an instrumental profile $\Psi(B)$ with a full-width at
half-maximum (FWHM) of $B$.

The optimal parameters $B, L, S$ and $O$ of the transformation are found by
minimising the $\chi^2$ between the Li\`ege and the modified Hamburg
spectrum with a Levenberg-Marquardt algorithm
\citepads{2009ASPC..411..251M}. The method directly yields the parameters of
the Li\`ege atlas relative to the Hamburg atlas. Under the assumption that
the characteristics of the Hamburg atlas are well known, the true spectral
resolution, wavelength stability, and continuum level of the Li\`ege
spectrum can be retrieved. Similar procedures are commonly employed to
estimate the instrumental profiles of spectrometers where a direct
measurement is not feasible \citepads[][]{1995PASP..107..966V}.

Both spectra were normalised to have their continuum at unity. The error
estimates for the spectra were computed by combining the nominal noise of
the Li\`ege spectrum and the effective noise of the modified Hamburg spectrum. The
S/N of the latter is increased during the optimisation process due to the
convolution with a kernel of non-zero width.

To minimise the impact of blending telluric lines, we applied the fitting
procedure to narrow wavelength intervals of $\pm 150$\,m\AA{} around
individual solar spectral lines. The line list provided by
\citetalias{Pierce+Breckinridge.1973} was used to select the target lines.
The overlapping spectral region of the Hamburg and Li\`ege atlases and the
line list is~3290--8770.7\,\AA. Of the total number of~9073 identified solar
lines in that range, we only processed those with a line depression of more
than~5\% relative to the local maximum intensity. To remove the most extreme
outliers from the~8035 processed lines, we discarded those with a reduced
$\chi^2$ ($\chi^2/n$, $n$: number of degrees of freedom of the fit) larger
than the 95th percentile of the initial $\chi^2/n$ distribution. This left a
total of~7634 fits that are analysed in the following.

An initial run was carried out with all four free parameters ($L$, $S$, $B$,
$O$) for the fits. From that test we convinced ourselfs that the straylight
parameter $O$ is indeed compatible with zero throughout the whole spectral
range (see discussion in the Appendix). The final fits were carried out with
only three free parameters ($L$, $S$, $B$).

\section{Results and discussion}
\label{sec:results}
\label{sec:discussion}

\begin{figure*}[tbp]
  \includegraphics[width=\textwidth]{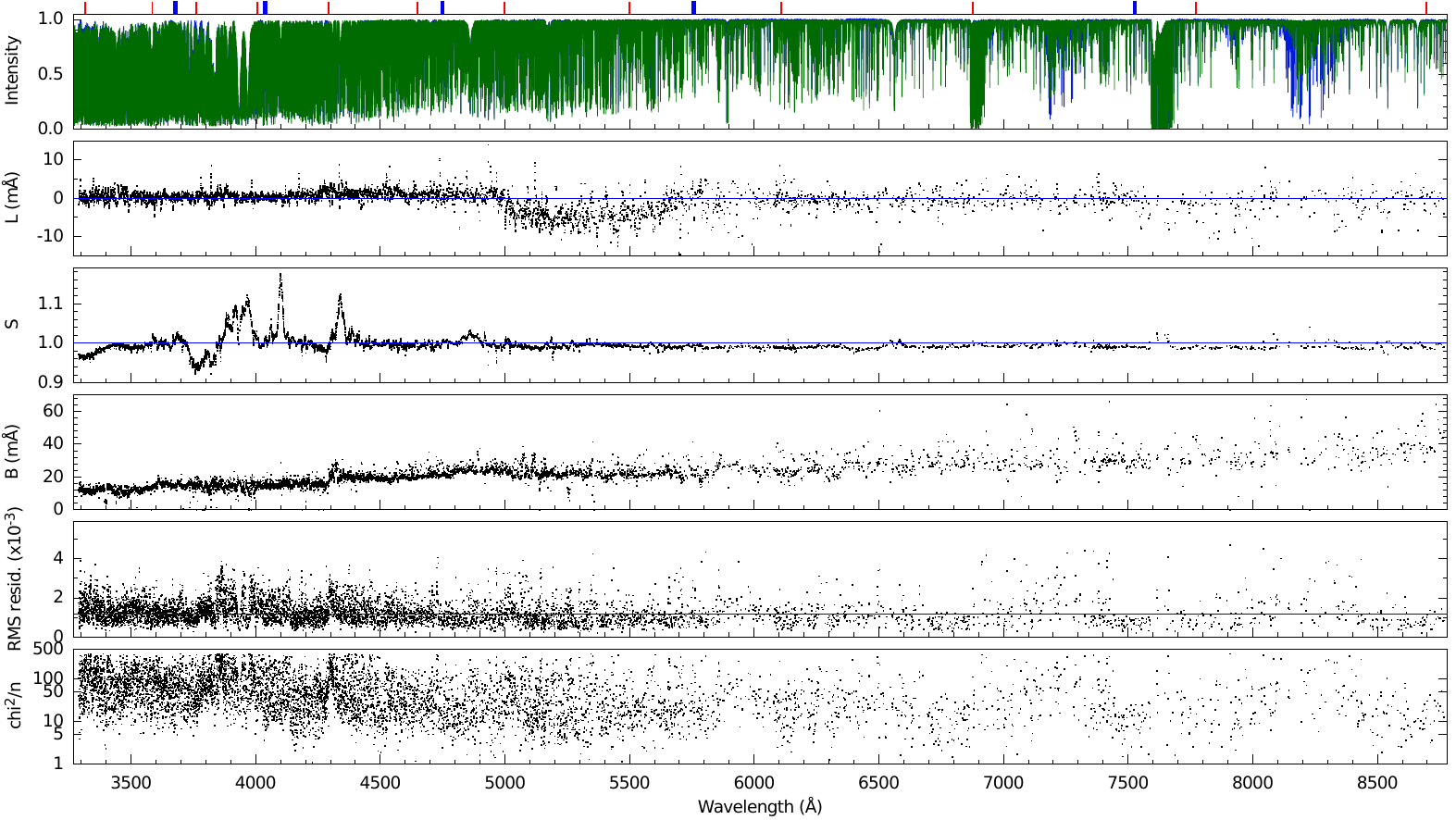}
  \caption{Investigated spectral region (topmost panel; blue trace:
    Hamburg, green trace: Li\`ege) and the results from the fitting
    procedure (parameters $L$, $S$, $B$ of
    Eq~(\ref{eq:transformation}), the RMS of the residuals of the
    fit, and the reduced $\chi^2$) versus wavelength. The thin red tick
    marks and the bold blue tick marks above the topmost panel indicate
    grating order intersections and FTS scan intersections,
    respectively. Zero wavelength shift and unity scaling factors are
    indicated by the horizontal blue lines in the $L$ and $S$ panels.  The
    blue line in the trace of the RMS residuals indicates the median
    value.}
  \label{fig:3par.masked.params}
\end{figure*}

Fig.~\ref{fig:3par.masked.params} summarises the results of all fits (see
Fig.~\ref{fig:example.fit} for an example of one individual fit). The
topmost panel displays a trace of the Hamburg and Li\`ege spectra in the
investigated wavelength interval. The $L$, $S$, and $B$ parameters are
plotted versus wavelength in panels two to four. In the lowest two panels,
the root mean square (RMS) of the residuals for each fit and the reduced
$\chi^2$ are plotted.
The individual parameters are discussed in the following subsections. The
results of the fits and derived correction parameters that can be applied to
e.g. synthesized spectra are available online at the Centre de Donn\'ees
astronomiques de Strasbourg~(CDS).

\subsection{Reliability of the fits}

The quality of the fits is very variable, but the parameters are within a
physically meaningful range. The median of $\chi^2/n$ is~42.4, the 95th
percentile is~386.8; 239 fits have a reduced $\chi^2$ better than~5. The
very good fits are essentially evenly distributed throughout the whole
wavelength range, but the $\chi^2$ improves with increasing wavelength in
general. This is because the line depths decrease towards the red part of
the spectrum which lowers the S/N of the line profiles.
\begin{figure}[tbp]
  \centering
  \includegraphics[width=6cm]{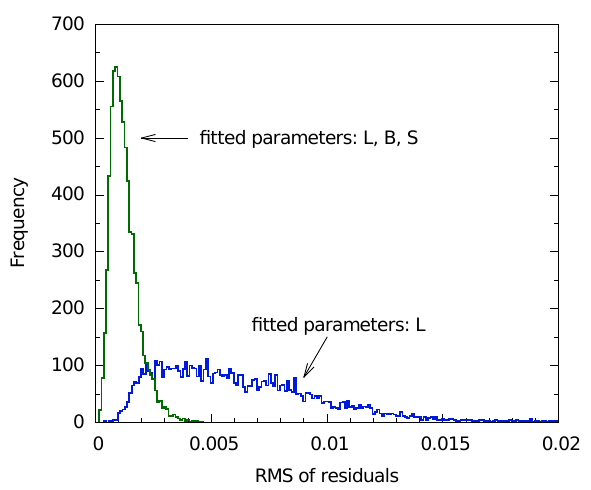}
  \caption{Histogram of the RMS of the fit residuals. The case when only the
    wavelength shifts are fitted is compared to the case where wavelength
    shift, scaling and broadening (panel~5 in
    Fig.~\ref{fig:3par.masked.params}) are considered. The width of the bins
    is $1\times10^{-4}$.}
  \label{fig:resid.histo}
\end{figure}

The $\chi^2/n$ values are always larger than unity which indicates that our
model does not account for---within the given very low noise levels---all
the differences between the two atlases. This can also be seen in the
residuals of an example fit shown in Fig.~\ref{fig:example.fit}. The
residuals show systematic variations on a scale narrower than telluric or
solar spectral features. This is a typical pattern and indicates that the
differences stem from instrumental artefacts (e.g. spectral fringes) or data
processing artefacts. Further effects that are not covered by our model but
that are most likely imprinted in the data to some degree are
non-linearities of the detectors (e.g. the discussion in
\citetalias{1984sfat.book.....K} about the impact on FTS spectra) and
intensity gradients within short wavelength intervals. After all we note
that the $\chi^2$ values computed here are crude estimates because we do not
have noise estimates at all wavelengths and because adjacent data points of
the Li\`ege spectrum are correlated to an unknown degree due to the heavy
post-processing of the original raw data (low-pass filtering, re-sampling).

A measure for the similarity of the spectra after applying the best-fitting
instrumental degradations is the RMS of the fit-residuals. The median of the
RMS residuals of all fits is~$1.18\times10^{-3}$, the 95th percentile
is~$2.5\times10^{-3}$. The histogram of the RMS residuals is shown in
Fig.~\ref{fig:resid.histo} in comparison to the case where only the $L$
parameter was fitted, that is, where only the wavelength shift between both
spectra was corrected. In the latter case, the distribution of the RMS
residuals is much broader and the median is about five times larger at
$5.7\times10^{-3}$.

\begin{figure}[tbp]
  \resizebox{\hsize}{!}{\includegraphics{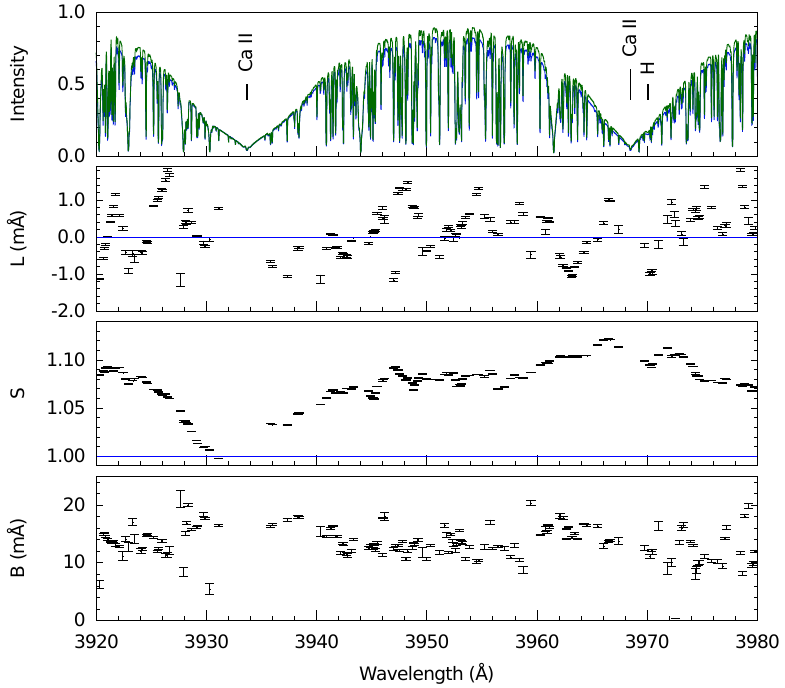}}
  \caption{Zoom to a narrow wavelength window in
    Fig.~\ref{fig:3par.masked.params}. The gap in the very core of the
    \ion{Ca}{ii}~3934\,\AA{} line is caused by the lack of strong enough
    spectral features. Error bars represent the 1-$\sigma$ parameter
    uncertainties provided by the minimisation routine.
    \label{fig:3par.masked.params.zoom}}
\end{figure}

The apparent scatter of the fit parameters that is visible in
Fig.~\ref{fig:3par.masked.params} is not at all entirely random but an
effect of the compressed display. Once zoomed to a range where the
individual data points can be
distinguished~(Fig.~\ref{fig:3par.masked.params.zoom}), the values of the
parameters are similar or follow a non-random, systematic pattern within
short wavelength intervals. The 1-$\sigma$ errors of the fit parameters are
considerably smaller than the observed variations in narrow wavelength
intervals, supporting the interpretation that these variations are indeed
caused by nonmodelled systematic effects in the atlases. The intersections
between the grating orders and FTS scans (indicated in
Fig.~\ref{fig:3par.masked.params}) leave no clear trace in the fitted
parameters.

It also shall be noted that our method can not necessarily discriminate
between instrumental effects and real differences of the spectra. Some
degree of realisation noise has to be present in both atlases, and in any
other measurement due to the finite temporal and spatial averaging. While
the Hamburg spectrum is pieced together from only seven individual
realisations of the average photospheric spectrum, the Li\`ege atlas is made
of thousands thereof in different stages of the solar cycle. But while the
resulting fluctuations could explain some of the scatter we see in the
fitted parameters, the statistics of the large sample of realisations makes
the mean values of our results more robust.

\subsubsection{Impact of telluric lines}

\begin{figure}[htbp]
  \resizebox{\hsize}{!}{\includegraphics{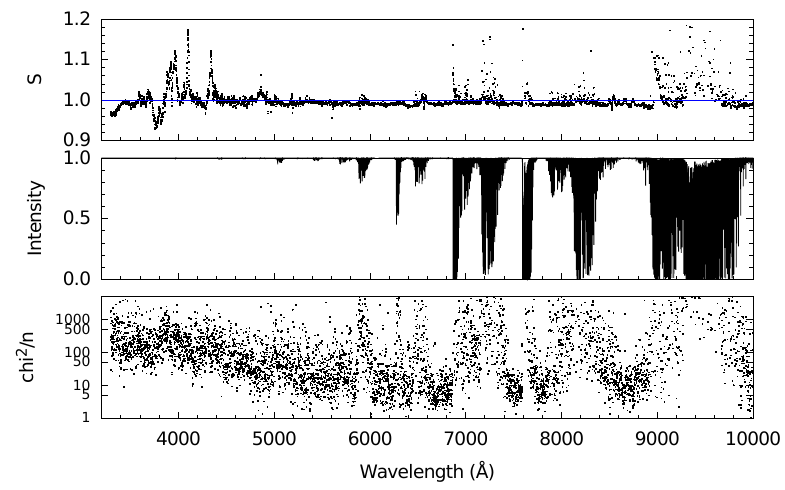}}
  \caption{Scaling parameter derived from fitting consecutive 1\,\AA{}
    intervals. A synthetic telluric spectrum is plotted for comparison with
    the pattern of the $\chi^2$ in the lowest panel.}
  \label{fig:3par.1A.params.tel}
\end{figure}

Telluric lines increasingly affect the fits redward of~5800\,\AA. This is
partly because telluric lines (especially H$_2$O lines) are stronger in the
Hamburg spectrum in general, but also because the different wavelength
shifts of the telluric lines with respect to the solar lines. The impact of
telluric lines is demonstrated in Fig.~\ref{fig:3par.1A.params.tel} where we
applied the fitting method to consecutive intervals of~1\,\AA{} width which
are therefore more susceptible to telluric blends. Shown are only the
parameter $S$, the reduced $\chi^2$ and a synthetic telluric spectrum which
was provided by \citet[][priv. comm.]{kurucz.tell.spect}. The $\chi^2/n$ is
strongly correlated with the telluric spectrum and to some degree also with
the $S$ parameter.

\subsection{Wavelength calibration}
\label{sec:wavelength.calib}

The wavelength calibrations of both atlases agree within the accuracy of the
used calibration reference when excluding the region~4900--5800\,\AA{}. The
mean of $L$ is $+0.5$\,m\AA, with a standard deviation of 1.8\,m\AA{} in
that region. This is not too surprising as both atlases are calibrated to
the same reference.

We do not have an explanation for the bump at~5000\,\AA{}, but it is clearly
due to a mis-calibration in the Li\`ege spectrum, not in the Hamburg
spectrum. We verified this by comparing measured positions of iron lines in
both atlases to the \citetalias{Pierce+Breckinridge.1973} line list and to
the respective laboratory wavelengths provided by
\citetads{1994ApJS...94..221N}. In both cases the same bump only appears for
the Li\`ege spectrum, not for the Hamburg spectrum. Interestingly,
\citetads{1998A&AS..129...41A} who used the same approach to compare the
Hamburg and the Li\`ege atlases, conclude that the Li\`ege spectrum is
affected by a substantial mis-calibration with a strong wavelength
dependence. A closer inspection revealed that these authors used a
particular version of the Li\`ege atlas that has a constant wavelength
offset of~14\,m\AA{} with respect to the official version used in this
work. When converted to a velocity, that offset fully explains the apparent
wavelength dependence observed by these authors.

In short wavelength intervals, the $L$ parameter sometimes follows
arc-shaped patterns. In the zoomed region around~3950\,\AA{} in
Fig.~\ref{fig:3par.masked.params.zoom}, the width of these arcs is
about~4-5\,\AA{} which is compatible with the width of the individually
observed segments at that wavelength. The pattern could be an artefact from
the non-linearity of the spectrograph's dispersion which might not have been
modelled in the wavelength scale of the individual segments. Furthermore,
\citetalias{Pierce+Breckinridge.1973} estimate that their systematic errors
can be as large as~2\,m\AA{}. Because of the fundamental differences in how
the Hamburg and Li\`ege atlas were pieced together from individual
observations, the systematic errors in the line list will to some degree
also show up in this analysis.

\subsection{Continuum level}
\label{sec:continuum.level}

The scaling parameter $S$ represents the ratio of the continuum intensity of
the Hamburg and Li\`ege spectra. At wavelengths larger
than~$\sim$\,5000\,\AA, $S$ is close to unity with a mean of~0.992 and a
standard deviation of~0.006. Large deviations from unity occur
below~$\sim$\,5000\,\AA{}, where we find several peaks between~0.94 and
1.175. Some of these peaks coincide with the Balmer lines (4861\,\AA,
4340\,\AA, 4101\,\AA, and 3970\,\AA) and the \ion{Ca}{ii} lines
at~3934\,\AA{} and~3968\,\AA{}
(cf.~Fig.~\ref{fig:3par.masked.params.zoom}). The departures from unity
farther to the blue seem not to be clearly coinciding with strong spectral
lines but are probably caused by the general lack of continuum in that
region. \citetalias{1984SoPh...90..205N} also provided a calibration for the
Li\`ege spectrum with respect to their absolute intensities and find a very
similar pattern (cf. their Fig.~12).

The $S$ parameter is less susceptible to telluric contamination than the $L$
and $B$ parameters. However, a comparison between
Fig.~\ref{fig:3par.1A.params.tel} and Fig.~\ref{fig:3par.masked.params}
suggests that some of the spikes in $S$ redward of 5500\,\AA{} are caused by
telluric contamination.

\citetalias{1973apds.book.....D} explicitly state that the continuum level
of the Li\`ege spectrum should not be trusted too much. Given the
manufacturing process of the Li\`ege atlas, it actually seems surprising
that the continuum level agrees so well with the Hamburg spectrum in most
spectral regions and emphasises their careful work.

\subsection{Line broadening and spectral resolution}
\label{sec:resolution}

The broadening parameter $B$ increases approximately linearly with
wavelength from $\sim$\,12\,m\AA{} to $\sim$\,40\,m\AA. It shows the
strongest scatter of the three fitted parameters. There is no clear and
pronounced systematic pattern visible on short wavelength intervals as it is
the case for the other parameters
(cf.~Fig~\ref{fig:3par.masked.params.zoom}). That scatter might be related
to the automatic adaptive low-pass filtering that was applied online during
data acquisition.

\begin{figure}[tbp]
  \resizebox{\hsize}{!}{\includegraphics{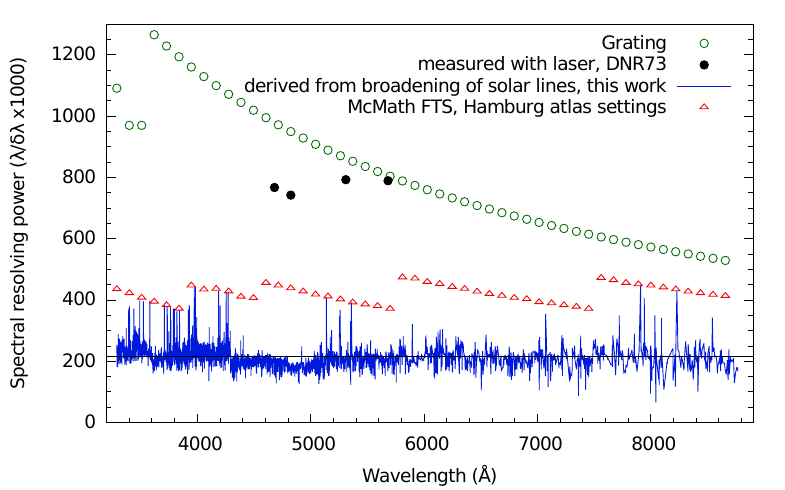}}
  \caption{Theoretical and measured resolving power of the Li\`ege
    atlas. The resolution settings for the individual segments of the
    Hamburg atlas are plotted for comparison. The plotted trace of the
    data measured in this work was filtered with a running median for
    better visibility. The horizontal black line marks the mean of our
    measured values.}
  \label{fig:jfjres}
\end{figure}

The large broadening that we observe throughout the whole Li\`ege spectrum
is in clear contrast to the stated spectral resolution.  From the fitted
parameter $B(\lambda)$ we derive the effective, absolute broadening of the
Li\`ege spectrum,
$B_{\text{Lieg}}(\lambda) = \left(W_{\text{FTS}}(\lambda)^2 +
  B(\lambda)^2\right)^{\frac 1 2}$,
by taking into account the known finite width $W_{\text{FTS}}(\lambda)$ of
the instrumental profile of the FTS. The spectral resolving power computed
from $R = \lambda/B_{\text{Lieg}}$ is plotted in Fig.~\ref{fig:jfjres} in
comparison to the theoretical performance of the gratings and the measured
resolving power. The laser measurements of \citetalias{1973apds.book.....D}
yield a nearly constant resolving power of $\sim$\,800\,000 which is only in
accordance with the theoretical resolution at~5682\,\AA{} (order 10). The
mean resolving power derived from $B_{\text{Lieg}}$ is almost constant with
wavelength with a mean of~216\,000. This is more than three times lower than
suggested by the laser measurements and between two to six times lower than
the theoretical grating performance.

\begin{figure}[tbp]
  \resizebox{\hsize}{!}{\includegraphics{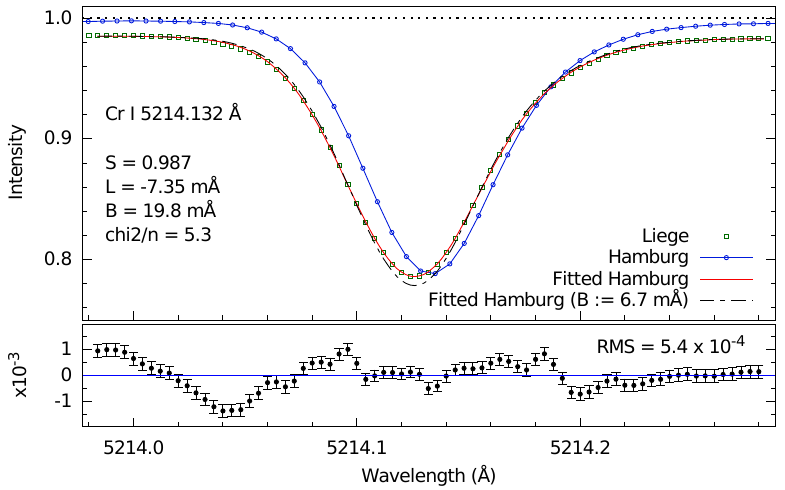}}
  \caption{Example of a fit to one spectral line. The blue dots show the
    original Hamburg spectrum, the red line shows the best-fitting modified
    Hamburg spectrum according to Equation~(\ref{eq:transformation}). The
    dashed line represents the fitted Hamburg spectrum but with the
    parameter $B$ forced to~6.7\,m\AA. The lower panel shows the residuals
    of the fit with three free parameters. Only each 2\textsuperscript{nd}
    data point is plotted in case of the Li\`ege spectrum for better
    visibility.}
  \label{fig:example.fit}
\end{figure}

We used only a simple Gaussian to model the broadening. However we carried
out tests with the instrumental profiles published in
\citetalias{1973apds.book.....D} which we scaled to the best-fitting width
and the results are virtually identical to a Gaussian of the same FWHM. As
the measured instrumental profiles are available only for four grating
orders, we decided to only use the more easily to reproduce gaussian model
here. We also need to stress that the published instrumental profiles can
not at all reproduce the observed line shapes. This is demonstrated in
Fig.~\ref{fig:example.fit} in case of a \ion{Cr}i line at 5214.132\,\AA. The
dashed black curve represents the modified Hamburg spectrum when $B$ is
fixed to~6.5\,m\AA{}, which is the FWHM of the published instrumental
profile for that wavelength, and only $S$ and $L$ are fitted. This leads to
a significant mismatch of the line shapes and the RMS of the fit residuals
increases to $2.7\times10^{-3}$ with a $\chi^2/n$ of~140. With three free
parameters however, the fit yields $B=19.8$\,m\AA{} with RMS residuals of
$5.4\times10^{-4}$ and a $\chi^2/n$ of~5.3.

We can only speculate about the origin of the reduction in spectral
resolution, but we can think of two effects that would at least
qualitatively explain the observed behaviour.  1) In case the solar image
happened to be aligned with the equator parallel to the slit the velocity
gradient of 800\,m\,s$^{-1}$ along the slit would translate to an
approximate broadening of~4\,m\AA{} at the blue end and~14\,m\AA{} at the
red end of the Li\`ege spectrum. 2) the so-called slit curvature (``spectrum
smile'') is an intrinsic distortion of a grating spectrograph which causes
the image of the entrance slit to be curved in the focal plane of the
spectrograph camera. With a scanning spectrograph, a spectral gradient is
imaged onto the exit slit which broadens the spectrum registered by the
detector. Laser measurements are less prone to the slit curvature in the
(likely) case that the slit is not fully illuminated by the laser. We do not
try to estimate the magnitude of that effect as it intimately depends on the
exact optical configuration of the spectrograph and the detector system.

\section{Conclusions}

Previous studies based on the Li\`ege and Hamburg spectral atlases sometimes
came to contradictory conclusions regarding the compatibility of parameters
retrieved from the both datasets.

We carried out a detailed comparison that takes into account the different
instrumental degradations (spectral broadening, wavelength shift, intensity
scaling, intensity offset) of both atlases. Our approach models these
degradations and tries to find the best-fitting model parameters to match
the Hamburg spectrum to the Li\`ege spectrum, thereby measuring Li\`ege's
wavelength calibration, continuum level, and resolution relative to the
Hamburg spectrum whose parameters are well known.
The results support the claimed quality metrics of the Li\`ege data except
for the spectral resolving power that we found to be between two to six
times lower than specified and which probably explains to a large extend the
disagreement of both atlases found by some authors. Studies based on
equivalent widths are susceptible to the position of the continuum but not
to the spectral resolution which however has an impact on model fitting.

But once known, the degradations can be accounted for. In doing so, we found
the RMS of the differences between both atlases to be less than
$1.2\times10^{-3}$ for 50\% of the~7634 investigated spectral lines in the
region between~3290 and~8771\,\AA{}. This is a remarkably good agreement
given the very different instruments and methods used to produce the two
atlases.

The Li\'ege and Hamburg atlases will probably continue to be the main solar
reference spectra for the coming years. Efforts are under way at the
astronomical institute of the university of G\"ottingen, Germany, to create
new, high-resolution, high-fidelity FTS atlases of the Sun
\citepads[][]{2016A&A...587A..65R}, but since the observations are carried
out from virtually sea level altitude, that data is expected to have an
enhanced telluric contamination compared to the Hamburg and Li\`ege
data. Thanks to the high-altitude site and the resulting lower contamination
with telluric H$_2$O features, the Li\`ege spectrum still is a
competitive---and in that regard also unique---option for many
investigations that rely on precise line shapes.

\begin{acknowledgements}
  The authors thank Robert L. Kurucz for providing his synthetic telluric
  spectrum, Ginette Roland for providing information about the observations
  at Jungfraujoch, Frank Hill for guiding us through the McMath FTS archive,
  and Ansgar Reiners for valuable comments on FT-spectrometers. We are
  grateful for the comments of the anonymous referee that helped to improve
  the manuscript. This work is partially supported by the Spanish Ministry
  of Science through projects AYA2014-55078-P, AYA2011-24808 and
  CSD2007-00050, and by the German Federal Ministry for Education and
  Research through the ``Role Of the MIddle atmosphere in Climate (ROMIC)''
  project.  This work contributes to the deliverables identified in FP7
  European Research Council grant agreement 277829, ``Magnetic connectivity
  through the Solar Partially Ionized Atmo- sphere''.  This research has
  made use of NASA's Astrophysics Data System.
\end{acknowledgements}

\bibliographystyle{aa}
\bibliography{references}

\begin{thebibliography}{46}
\expandafter\ifx\csname natexlab\endcsname\relax\def\natexlab#1{#1}\fi

\bibitem[{{Allende Prieto} {et~al.}(2004){Allende Prieto}, {Asplund}, \&
  {Fabiani Bendicho}}]{2004A&A...423.1109A}
{Allende Prieto}, C., {Asplund}, M., \& {Fabiani Bendicho}, P. 2004, \aap, 423,
  1109

\bibitem[{{Allende Prieto} \& {Garc{\'i}a
  L{\'o}pez}(1998)}]{1998A&AS..129...41A}
{Allende Prieto}, C. \& {Garc{\'i}a L{\'o}pez}, R.~J. 1998, \aaps, 129, 41

\bibitem[{{Ardeberg} \& {Virdefors}(1975)}]{1975A&A....45...19A}
{Ardeberg}, A. \& {Virdefors}, B. 1975, \aap, 45, 19

\bibitem[{{Asplund} {et~al.}(2009){Asplund}, {Grevesse}, {Sauval}, \&
  {Scott}}]{2009ARA&A..47..481A}
{Asplund}, M., {Grevesse}, N., {Sauval}, A.~J., \& {Scott}, P. 2009, \araa, 47,
  481

\bibitem[{{Asplund} {et~al.}(2000){Asplund}, {Nordlund}, {Trampedach}, {Allende
  Prieto}, \& {Stein}}]{2000A&A...359..729A}
{Asplund}, M., {Nordlund}, {\AA}., {Trampedach}, R., {Allende Prieto}, C., \&
  {Stein}, R.~F. 2000, \aap, 359, 729

\bibitem[{{Ayres}(2008)}]{2008ApJ...686..731A}
{Ayres}, T.~R. 2008, \apj, 686, 731

\bibitem[{{Balthasar}(1988)}]{1988A&AS...72..473B}
{Balthasar}, H. 1988, \aaps, 72, 473

\bibitem[{{Beck} {et~al.}(2011){Beck}, {Rezaei}, \&
  {Fabbian}}]{2011A&A...535A.129B}
{Beck}, C., {Rezaei}, R., \& {Fabbian}, D. 2011, \aap, 535, A129

\bibitem[{{Beckers} {et~al.}(1976){Beckers}, {Bridges}, \&
  {Gilliam}}]{1976hrsa.book.....B}
{Beckers}, J.~M., {Bridges}, C.~A., \& {Gilliam}, L.~B. 1976, {A high
  resolution spectral atlas of the solar irradiance from 380 to 700 nanometers.
  Volume 2: Graphical form}

\bibitem[{{Brault}(1985)}]{1985hra..conf....3B}
{Brault}, J.~W. 1985, in High Resolution in Astronomy, Fifteenth Advanced
  Course of the Swiss Society of Astronomy and Astrophysics. Edited by A.O.
  Benz, M. Huber, and M. Mayer. Sauverny, Switzerland : Geneva Observatory,
  1985, p. 3-61, ed. A.~O. {Benz}, M.~{Huber}, \& M.~{Mayor}, 3--61

\bibitem[{{Caffau} {et~al.}(2008){Caffau}, {Ludwig}, {Steffen}, {Ayres},
  {Bonifacio}, {Cayrel}, {Freytag}, \& {Plez}}]{2008A&A...488.1031C}
{Caffau}, E., {Ludwig}, H.-G., {Steffen}, M., {et~al.} 2008, \aap, 488, 1031

\bibitem[{{Caffau} {et~al.}(2011){Caffau}, {Ludwig}, {Steffen}, {Freytag}, \&
  {Bonifacio}}]{2011SoPh..268..255C}
{Caffau}, E., {Ludwig}, H.-G., {Steffen}, M., {Freytag}, B., \& {Bonifacio}, P.
  2011, \solphys, 268, 255

\bibitem[{{Caffau} {et~al.}(2015){Caffau}, {Ludwig}, {Steffen}, {Livingston},
  {Bonifacio}, {Malherbe}, {Doerr}, \& {Schmidt}}]{2015A&A...579A..88C}
{Caffau}, E., {Ludwig}, H.-G., {Steffen}, M., {et~al.} 2015, \aap, 579, A88

\bibitem[{{Caffau} {et~al.}(2009){Caffau}, {Maiorca}, {Bonifacio},
  {Faraggiana}, {Steffen}, {Ludwig}, {Kamp}, \& {Busso}}]{2009A&A...498..877C}
{Caffau}, E., {Maiorca}, E., {Bonifacio}, P., {et~al.} 2009, \aap, 498, 877

\bibitem[{{Delbouille} \& {Roland}(1995)}]{1995ASPC...81...32D}
{Delbouille}, L. \& {Roland}, C. 1995, in Astronomical Society of the Pacific
  Conference Series, Vol.~81, Laboratory and Astronomical High Resolution
  Spectra, ed. A.~J. {Sauval}, R.~{Blomme}, \& N.~{Grevesse}, 32

\bibitem[{{Delbouille} \& {Roland}(1963)}]{1963apss.book.....D}
{Delbouille}, L. \& {Roland}, G. 1963, {Atlas photometrique du spectre solaire
  de $\lambda$\,7498 a $\lambda$\,12016. Photometric atlas of the solar
  spectrum from $\lambda$\,7498 to $\lambda$\,12016. [Par] L. Delbouille [et]
  G. Roland.}

\bibitem[{{Delbouille} {et~al.}(1973){Delbouille}, {Roland}, \&
  {Neven}}]{1973apds.book.....D}
{Delbouille}, L., {Roland}, G., \& {Neven}, L. 1973, {Atlas photometrique du
  spectre solaire de $\lambda$\,3000 a $\lambda$\,10000}

\bibitem[{{Fabbian} \& {Moreno-Insertis}(2015)}]{2015ApJ...802...96F}
{Fabbian}, D. \& {Moreno-Insertis}, F. 2015, \apj, 802, 96

\bibitem[{{Gurtovenko} \& {Kostik}(1982)}]{1982A&AS...47..193G}
{Gurtovenko}, E.~A. \& {Kostik}, R.~I. 1982, \aaps, 47, 193

\bibitem[{{Hinkle} {et~al.}(1995){Hinkle}, {Wallace}, \&
  {Livingston}}]{1995ASPC...81...66H}
{Hinkle}, K.~H., {Wallace}, L., \& {Livingston}, W. 1995, in Astronomical
  Society of the Pacific Conference Series, Vol.~81, Laboratory and
  Astronomical High Resolution Spectra, ed. A.~J. {Sauval}, R.~{Blomme}, \&
  N.~{Grevesse}, 66

\bibitem[{{Kiselman}(1994)}]{1994A&AS..104...23K}
{Kiselman}, D. 1994, \aaps, 104

\bibitem[{{Kurucz}(2005)}]{2005MSAIS...8..189K}
{Kurucz}, R.~L. 2005, Memorie della Societa Astronomica Italiana Supplementi,
  8, 189

\bibitem[{{Kurucz}(2015)}]{kurucz.tell.spect}
{Kurucz}, R.~L. 2015, private communication

\bibitem[{{Kurucz} {et~al.}(1984){Kurucz}, {Furenlid}, {Brault}, \&
  {Testerman}}]{1984sfat.book.....K}
{Kurucz}, R.~L., {Furenlid}, I., {Brault}, J., \& {Testerman}, L. 1984, {Solar
  flux atlas from 296 to 1300 nm}

\bibitem[{{Labs} \& {Neckel}(1967)}]{1967ZA.....65..133L}
{Labs}, D. \& {Neckel}, H. 1967, \zap, 65, 133

\bibitem[{{Livingston} \& {Wallace}(1987)}]{1987ApJ...314..808L}
{Livingston}, W. \& {Wallace}, L. 1987, \apj, 314, 808

\bibitem[{{Livingston} {et~al.}(2007){Livingston}, {Wallace}, {White}, \&
  {Giampapa}}]{2007ApJ...657.1137L}
{Livingston}, W., {Wallace}, L., {White}, O.~R., \& {Giampapa}, M.~S. 2007,
  \apj, 657, 1137

\bibitem[{{Livingston} {et~al.}(1988){Livingston}, {Wallace}, \&
  {White}}]{1988Sci...240.1765L}
{Livingston}, W.~C., {Wallace}, L., \& {White}, O.~R. 1988, Science, 240, 1765

\bibitem[{{Markwardt}(2009)}]{2009ASPC..411..251M}
{Markwardt}, C.~B. 2009, in Astronomical Society of the Pacific Conference
  Series, Vol. 411, Astronomical Data Analysis Software and Systems XVIII, ed.
  D.~A. {Bohlender}, D.~{Durand}, \& P.~{Dowler}, 251

\bibitem[{{Minnaert} {et~al.}(1940){Minnaert}, {Houtgast}, \&
  {Mulders}}]{1940pass.book.....M}
{Minnaert}, M., {Houtgast}, J., \& {Mulders}, G.~F.~W. 1940, {Photometric atlas
  of the solar spectrum from [lambda] 3612 to [lambda] 8771 with an appendix
  from [lambda] 3332 to [lambda] 3637}

\bibitem[{{Nave} {et~al.}(1994){Nave}, {Johansson}, {Learner}, {Thorne}, \&
  {Brault}}]{1994ApJS...94..221N}
{Nave}, G., {Johansson}, S., {Learner}, R.~C.~M., {Thorne}, A.~P., \& {Brault},
  J.~W. 1994, \apjs, 94, 221

\bibitem[{{Neckel}(1999)}]{1999SoPh..184..421N}
{Neckel}, H. 1999, \solphys, 184, 421

\bibitem[{{Neckel} \& {Labs}(1984)}]{1984SoPh...90..205N}
{Neckel}, H. \& {Labs}, D. 1984, \solphys, 90, 205

\bibitem[{{Pereira} {et~al.}(2013){Pereira}, {Asplund}, {Collet}, {Thaler},
  {Trampedach}, \& {Leenaarts}}]{2013A&A...554A.118P}
{Pereira}, T.~M.~D., {Asplund}, M., {Collet}, R., {et~al.} 2013, \aap, 554,
  A118

\bibitem[{{Pereira} {et~al.}(2009){Pereira}, {Kiselman}, \&
  {Asplund}}]{2009A&A...507..417P}
{Pereira}, T.~M.~D., {Kiselman}, D., \& {Asplund}, M. 2009, \aap, 507, 417

\bibitem[{{Pierce} \& {Breckinridge}(1973)}]{Pierce+Breckinridge.1973}
{Pierce}, A. \& {Breckinridge}, J. 1973, The Kitt-Peak Table of Solar Spectrum
  Wavelengths, Vol. Contribution No. 559 (Kitt Peak National Observatory)

\bibitem[{{Reiners} {et~al.}(2016){Reiners}, {Mrotzek}, {Lemke}, {Hinrichs}, \&
  {Reinsch}}]{2016A&A...587A..65R}
{Reiners}, A., {Mrotzek}, N., {Lemke}, U., {Hinrichs}, J., \& {Reinsch}, K.
  2016, \aap, 587, A65

\bibitem[{{Rutten} \& {van der Zalm}(1984)}]{1984A&AS...55..143R}
{Rutten}, R.~J. \& {van der Zalm}, E.~B.~J. 1984, \aaps, 55, 143

\bibitem[{{Scott} {et~al.}(2015){Scott}, {Asplund}, {Grevesse}, {Bergemann}, \&
  {Sauval}}]{2015A&A...573A..26S}
{Scott}, P., {Asplund}, M., {Grevesse}, N., {Bergemann}, M., \& {Sauval}, A.~J.
  2015, \aap, 573, A26

\bibitem[{{Stenflo}(2015{\natexlab{a}})}]{2015A&A...573A..74S}
{Stenflo}, J.~O. 2015{\natexlab{a}}, \aap, 573, A74

\bibitem[{{Stenflo} {et~al.}(1983){Stenflo}, {Twerenbold}, {Harvey}, \&
  {Brault}}]{1983A&AS...54..505S}
{Stenflo}, J.~O., {Twerenbold}, D., {Harvey}, J.~W., \& {Brault}, J.~W. 1983,
  \aaps, 54, 505

\bibitem[{{Stenflo}(2015{\natexlab{b}})}]{sten2015pc}
{Stenflo}, O. 2015{\natexlab{b}}, private communication

\bibitem[{{Valenti} {et~al.}(1995){Valenti}, {Butler}, \&
  {Marcy}}]{1995PASP..107..966V}
{Valenti}, J.~A., {Butler}, R.~P., \& {Marcy}, G.~W. 1995, \pasp, 107, 966

\bibitem[{{Wallace} {et~al.}(1998){Wallace}, {Hinkle}, \&
  {Livingston}}]{1998assp.book.....W}
{Wallace}, L., {Hinkle}, K., \& {Livingston}, W. 1998, {An atlas of the
  spectrum of the solar photosphere from 13,500 to 28,000 cm$^{-1}$ (3570 to
  7405 \AA{})} (National Solar Observatory)

\bibitem[{{Wallace} {et~al.}(2007){Wallace}, {Hinkle}, \&
  {Livingston}}]{2007assp.book.....W}
{Wallace}, L., {Hinkle}, K., \& {Livingston}, W. 2007, {An Atlas of the
  Spectrum of the Solar Photosphere from 13,500 to 33,980 cm$^{-1}$ (2942 to
  7405 {\AA})}

\bibitem[{{Wallace} {et~al.}(2011){Wallace}, {Hinkle}, {Livingston}, \&
  {Davis}}]{2011ApJS..195....6W}
{Wallace}, L., {Hinkle}, K.~H., {Livingston}, W.~C., \& {Davis}, S.~P. 2011,
  \apjs, 195, 6

\end{thebibliography}

\begin{appendix}
\label{sec:appendix}

\section{Spectral scattered light}
\label{sec:scattered.light}

\begin{figure}[htbp]
  \resizebox{\hsize}{!}{\includegraphics{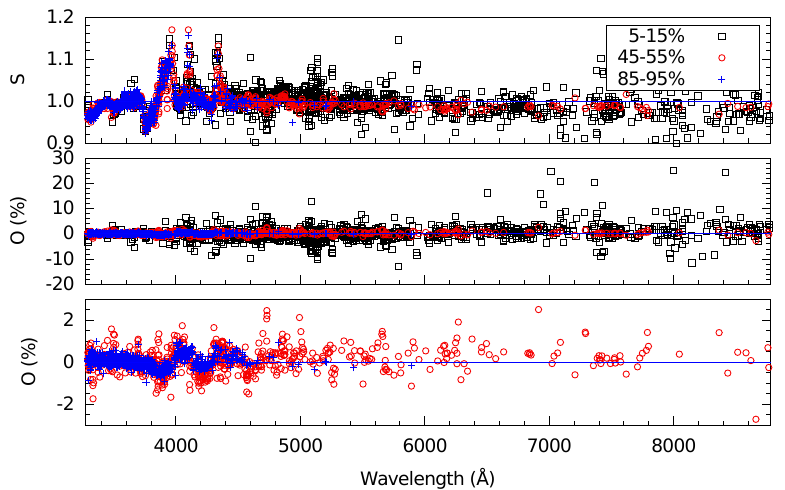}}\\
  \resizebox{\hsize}{!}{\includegraphics{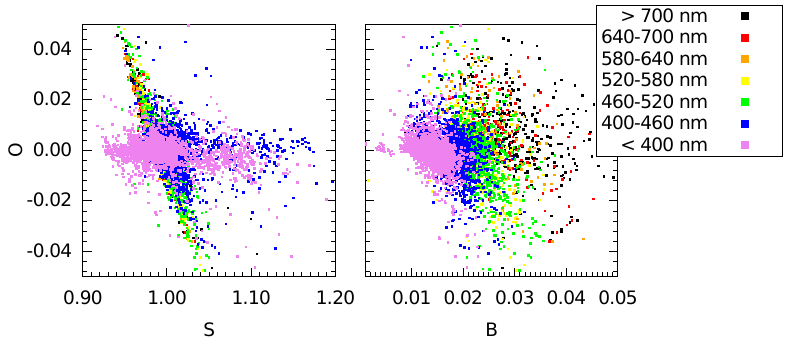}}
  \caption{Correlation between the scale and offset
    parameters. \textit{Upper panel}: Scale and offset parameters for
    lines with three different regimes of line depths. \textit{Lower
      panel}: Scatter plots of the offset parameter vs. the scale and
    broadening parameters. The colours represent different spectral
    intervals to distinguish chromatic effects.}
  \label{fig:4par.offset.scale}
\end{figure}

The parameter $O$ in Equation~(\ref{eq:transformation}) represents a
constant offset as it can be caused by spectral straylight. Here we show
that the amount of scattered light in the above definition is virtually zero
and can be neglected in the analysis. The upper panel of
Fig.~\ref{fig:4par.offset.scale} shows the parameters $S$ and $O$ plotted
vs. wavelength for lines with three different regimes of the line depth
(5-15\%, 45-55\%, 85-95\%).
While the mean of $O$ is below 0.1\% in all cases, the scatter in $S$ and
$O$ decreases dramatically with increasing line depth with standard
deviations of 3.2\%, 0.6\% and 0.26\% for the line depths given above. This
dependence on the line depth is expected because only strong intensity
variations within the fitted spectral range can simultaneously constrain the
$O$ and $S$ parameters.

The strong anti-correlation between $O$ and $S$ becomes also apparent in the
scatter plot in the lower panel of Fig.~\ref{fig:4par.offset.scale}.  The
fit tries to compensate a higher continuum level with negative offset
parameters and vice versa, except for wavelengths below~400\,nm where the
spectrum is dominated by deep lines
(cf. Fig.~\ref{fig:3par.masked.params}). $B$ and $O$ are essentially
uncorrelated except for a chromatic effect caused by the wavelength
dependence of $B$.

We therefore can confirm the stated excellent straylight suppression of the
Jungfraujoch spectrograph and it is justified to enforce $O$ to zero for our
analysis. Limiting the optimisation to three free parameters stabilises the
fits and the scatter in the~$S$ parameter is greatly reduced with only a
slight increase of the~$\chi^2$ values.

\end{appendix}

\end{document}